\documentclass[10pt]{IEEEtran}
\usepackage{url}
\usepackage[utf8]{inputenc}
\usepackage{color}
\usepackage{amsmath}
\usepackage{amssymb}

\usepackage[acronyms,nonumberlist,nopostdot,nomain,nogroupskip]{glossaries}
\usepackage{tablefootnote}
\usepackage{booktabs}

\usepackage{tikz}
\usepackage{pgfplots}
\pgfplotsset{compat=newest} 
\pgfplotsset{plot coordinates/math parser=false} 
\newlength\fheight
\newlength\fwidth
\usetikzlibrary{plotmarks,patterns,decorations.pathreplacing,backgrounds,calc,arrows,arrows.meta,spy,matrix}
\usepgfplotslibrary{patchplots,groupplots}
\usepackage{tikzscale}
\usepackage{hyperref}

\usepackage{multirow}
\usepackage{tkz-kiviat}

\usepackage[font=scriptsize]{subcaption}
\usepackage[font=footnotesize]{caption}
\usepackage{float}

\usepackage{mathtools}

\newacronym{3gpp}{3GPP}{3rd Generation Partnership Project}
\newacronym{adc}{ADC}{Analog to Digital Converter}
\newacronym{5g}{5G}{5th generation}
\newacronym{aimd}{AIMD}{Additive Increase Multiplicative Decrease}
\newacronym{am}{AM}{Acknowledged Mode}
\newacronym{amc}{AMC}{Adaptive Modulation and Coding}
\newacronym{aqm}{AQM}{Active Queue Management}
\newacronym{awgn}{AGWN}{Additive White Gaussian Noise}
\newacronym{balia}{BALIA}{Balanced Link Adaptation}
\newacronym{bdp}{BDP}{Bandwidth-Delay Product}
\newacronym{bf}{BF}{Beamforming}
\newacronym{cc}{CC}{Congestion Control}
\newacronym{cdf}{CDF}{Cumulative Distribution Function}
\newacronym{cn}{CN}{Core Network}
\newacronym{cqi}{CQI}{Channel Quality Information}
\newacronym{cp}{CP}{Control Plane}
\newacronym{csirs}{CSI-RS}{Channel State Information - Reference Signal}
\newacronym{dc}{DC}{Dual Connectivity}
\newacronym{dce}{DCE}{Direct Code Execution}
\newacronym{dci}{DCI}{Downlink Control Information}
\newacronym{dl}{DL}{Downlink}
\newacronym{dmr}{DMR}{Deadline Miss Ratio}
\newacronym{dmrs}{DMRS}{DeModulation Reference Signal}
\newacronym{e2e}{E2E}{End-to-End}
\newacronym{ecn}{ECN}{Explicit Congestion Notification}
\newacronym{edf}{EDF}{Earliest Deadline First}
\newacronym{enb}{eNB}{evolved Node Base}
\newacronym{epc}{EPC}{Evolved Packet Core}
\newacronym{es}{ES}{Edge Server}
\newacronym{fdma}{FDMA}{Frequency Division Multiple Access}
\newacronym{fdd}{FDD}{Frequency Division Duplexing}
\newacronym[firstplural=Radio Access Technologies (RATs)]{rat}{RAT}{Radio Access Technology}
\newacronym{fs}{FS}{Fast Switching}
\newacronym{ftp}{FTP}{File Transfer Protocol}
\newacronym{gnb}{gNB}{Next Generation Node Base Station}
\newacronym{harq}{HARQ}{Hybrid Automatic Repeat reQuest}
\newacronym{hetnet}{HetNet}{Heterogeneous Network}
\newacronym{hh}{HH}{Hard Handover}
\newacronym{hol}{HOL}{Head-of-Line}
\newacronym{ia}{IA}{Initial Access}
\newacronym{imt}{IMT}{International Mobile Telecommunication}
\newacronym{iot}{IoT}{Internet of Things}
\newacronym{los}{LOS}{Line of Sight}
\newacronym{lte}{LTE}{Long Term Evolution}
\newacronym{m2m}{M2M}{Machine to Machine}
\newacronym{mac}{MAC}{Medium Access Control}
\newacronym{mc}{MC}{Multi-Connectivity}
\newacronym{mcs}{MCS}{Modulation and Coding Scheme}
\newacronym{mec}{MEC}{Mobile Edge Cloud}
\newacronym{mi}{MI}{Mutual Information}
\newacronym{mimo}{MIMO}{Multiple Input, Multiple Output}
\newacronym{mmwave}{mmWave}{millimeter wave}
\newacronym{mptcp}{MPTCP}{Multipath TCP}
\newacronym{mr}{MR}{Maximum Rate}
\newacronym{mss}{MSS}{Maximum Segment Size}
\newacronym{mtd}{MTD}{Machine-Type Device}
\newacronym{mtu}{MTU}{Maximum Transmission Unit}
\newacronym{nfv}{NFV}{Network Function Virtualization}
\newacronym{nlos}{NLOS}{Non Line of Sight}
\newacronym{nr}{NR}{NR}
\newacronym{ofdm}{OFDM}{Orthogonal Frequency Division Multiplexing}
\newacronym{pdcch}{PDCCH}{Physical Downlonk Control Channel}
\newacronym{pdcp}{PDCP}{Packet Data Convergence Protocol}
\newacronym{pdsch}{PDSCH}{Physical Downlink Shared Channel}
\newacronym{pdu}{PDU}{Packet Data Unit}
\newacronym{pf}{PF}{Proportional Fair}
\newacronym{pgw}{PGW}{Packet Gateway}
\newacronym{phy}{PHY}{Physical}
\newacronym{pbch}{PBCH}{Physical Broadcast Channel}
\newacronym[plural=\gls{mme}s,firstplural=Mobility Management Entities (MMEs)]{mme}{MME}{Mobility Management Entity}
\newacronym{prb}{PRB}{Physical Resource Block}
\newacronym{pss}{PSS}{Primary Synchronization Signal}
\newacronym{pucch}{PUCCH}{Physical Uplink Control Channel}
\newacronym{pusch}{PUSCH}{Physical Uplink Shared Channel}
\newacronym{rach}{RACH}{Random Access Channel}
\newacronym{ran}{RAN}{Radio Access Network}
\newacronym{red}{RED}{Random Early Detection}
\newacronym{rf}{RF}{Radio Frequency}
\newacronym{rlc}{RLC}{Radio Link Control}
\newacronym{rlf}{RLF}{Radio Link Failure}
\newacronym{rrc}{RRC}{Radio Resource Control}
\newacronym{rrm}{RRM}{Radio Resource Management}
\newacronym{rr}{RR}{Round Robin}
\newacronym{rs}{RS}{Remote Server}
\newacronym{rsrp}{RSRP}{Reference Signal Received Power}
\newacronym{rss}{RSS}{Received Signal Strength}
\newacronym{rtt}{RTT}{Round Trip Time}
\newacronym{rw}{RW}{Receive Window}
\newacronym{rx}{RX}{Receiver}
\newacronym{sa}{SA}{Standalone}
\newacronym{sack}{SACK}{Selective Acknowledgment}
\newacronym{sap}{SAP}{Service Access Point}
\newacronym{sch}{SCH}{Secondary Cell Handover}
\newacronym{scoot}{SCOOT}{Split Cycle Offset Optimization Technique}
\newacronym{sdma}{SDMA}{Spatial Division Multiple Access}
\newacronym{sinr}{SINR}{Signal to Interference plus Noise Ratio}
\newacronym{sm}{SM}{Saturation Mode}
\newacronym{snr}{SNR}{Signal to Noise Ratio}
\newacronym{son}{SON}{Self-Organizing Network}
\newacronym{ss}{SS}{Synchronization Signal}
\newacronym{srs}{SRS}{Sounding Reference Signal}
\newacronym{sss}{SSS}{Secondary Synchronization Signal}
\newacronym{tb}{TB}{Transport Block}
\newacronym{tcp}{TCP}{Transmission Control Protocol}
\newacronym{tdd}{TDD}{Time Division Duplexing}
\newacronym{tdma}{TDMA}{Time Division Multiple Access}
\newacronym{tfl}{TfL}{Transport for London}
\newacronym{tm}{TM}{Transparent Mode}
\newacronym{trp}{TRP}{Transmitter Receiver Pair}
\newacronym{tti}{TTI}{Transmission Time Interval}
\newacronym{ttt}{TTT}{Time-to-Trigger}
\newacronym{tx}{TX}{Transmitter}
\newacronym{ue}{UE}{User Equipment}
\newacronym{ul}{UL}{Uplink}
\newacronym{uml}{UML}{Unified Modeling Language}
\newacronym{um}{UM}{Unacknowledged Mode}
\newacronym{utc}{UTC}{Urban Traffic Control}
\newacronym{vm}{VM}{Virtual Machine}
\newacronym{rsrq}{RSRQ}{Reference Signal Received Quality}
\newacronym{rssi}{RSSI}{Received Signal Strength Indicator}
\newacronym{crs}{CRS}{Cell Reference Signal}
\newacronym{nsa}{NSA}{Non-Standalone}
\newacronym{mrdc}{MR-DC}{Multi \gls{rat} \gls{dc}}
\newacronym{endc}{EN-DC}{E-UTRAN-\gls{nr} \gls{dc}}
\newacronym{5gc}{5GC}{5G Core}
\tikzstyle{startstop} = [rectangle, rounded corners, minimum width=2cm, minimum height=0.5cm,text centered, draw=black]
\tikzstyle{io} = [trapezium, trapezium left angle=70, trapezium right angle=110, minimum width=3cm, minimum height=1cm, text centered, draw=black]
\tikzstyle{process} = [rectangle, minimum width=2cm, minimum height=0.5cm, text centered, draw=black, alignb=center]
\tikzstyle{decision} = [ellipse, minimum width=2cm, minimum height=1cm, text centered, draw=black]
\tikzstyle{arrow} = [thick,<->,>=stealth]
\tikzstyle{line} = [thick,>=stealth]
\tikzstyle{darrow} = [thick,<->,>=stealth,dashed]
\tikzstyle{sarrow} = [thick,->,>=stealth]
\tikzstyle{larrow} = [line width=0.1mm,dashdotted,->,>=stealth]

\makeatletter
\def\grd@save@target#1{%
  \def\grd@target{#1}}
\def\grd@save@start#1{%
  \def\grd@start{#1}}
\tikzset{
  grid with coordinates/.style={
    to path={%
      \pgfextra{%
        \edef\grd@@target{(\tikztotarget)}%
        \tikz@scan@one@point\grd@save@target\grd@@target\relax
        \edef\grd@@start{(\tikztostart)}%
        \tikz@scan@one@point\grd@save@start\grd@@start\relax
        \draw[minor help lines] (\tikztostart) grid (\tikztotarget);
        \draw[major help lines] (\tikztostart) grid (\tikztotarget);
        \grd@start
        \pgfmathsetmacro{\grd@xa}{\the\pgf@x/1cm}
        \pgfmathsetmacro{\grd@ya}{\the\pgf@y/1cm}
        \grd@target
        \pgfmathsetmacro{\grd@xb}{\the\pgf@x/1cm}
        \pgfmathsetmacro{\grd@yb}{\the\pgf@y/1cm}
        \pgfmathsetmacro{\grd@xc}{\grd@xa + \pgfkeysvalueof{/tikz/grid with coordinates/major step x}}
        \pgfmathsetmacro{\grd@yc}{\grd@ya + \pgfkeysvalueof{/tikz/grid with coordinates/major step y}}
        \foreach \x in {\grd@xa,\grd@xc,...,\grd@xb}
        \node[anchor=north] at (\x,\grd@ya) {\pgfmathprintnumber{\x}};
        \foreach \y in {\grd@ya,\grd@yc,...,\grd@yb}
        \node[anchor=east] at (\grd@xa,\y) {\pgfmathprintnumber{\y}};
      }
    }
  },
  minor help lines/.style={
    help lines,
    gray,
    line cap =round,
    xstep=\pgfkeysvalueof{/tikz/grid with coordinates/minor step x},
    ystep=\pgfkeysvalueof{/tikz/grid with coordinates/minor step y}
  },
  major help lines/.style={
    help lines,
    line cap =round,
    line width=\pgfkeysvalueof{/tikz/grid with coordinates/major line width},
    xstep=\pgfkeysvalueof{/tikz/grid with coordinates/major step x},
    ystep=\pgfkeysvalueof{/tikz/grid with coordinates/major step y}
  },
  grid with coordinates/.cd,
  minor step x/.initial=.5,
  minor step y/.initial=.2,
  major step x/.initial=1,
  major step y/.initial=1,
  major line width/.initial=1pt,
}
\makeatother

\makeglossaries

\begin{document}

\flushbottom
	
\glsunset{nr}
\glsunset{3gpp}

\title{Standalone and Non-Standalone\\Beam Management for 3GPP NR at mmWaves}

\author{\IEEEauthorblockN{Marco Giordani, \IEEEmembership{Student Member, IEEE}, Michele Polese, \IEEEmembership{Student Member, IEEE},  Arnab Roy, \IEEEmembership{Member, IEEE}, Douglas Castor, \IEEEmembership{Member, IEEE}, Michele Zorzi, \IEEEmembership{Fellow, IEEE}}
\thanks{Marco Giordani, Michele Polese and Michele Zorzi are with the Department of Information Engineering (DEI), University of Padova, Italy, and Consorzio Futuro in Ricerca (CFR), Italy.
Email:\{giordani,polesemi,zorzi\}@dei.unipd.it.

Arnab Roy and Douglas Castor are with InterDigital Communications, Inc., USA. 
Email:\{arnab.roy,douglas.castor\}@interdigital.com.
}
}

\maketitle

\begin{abstract}
The next generation of cellular networks will exploit mmWave frequencies to dramatically increase the network capacity. The communication at such high frequencies, however, requires directionality to compensate the increase in propagation loss. Users and base stations need to align their beams  during both initial access and data transmissions, to ensure the maximum gain is reached. The accuracy of the beam selection, and the delay in updating the beam pair or performing initial access, impact the end-to-end performance and the quality of service. In this paper we will present the beam management procedures that 3GPP has included in the NR specifications, focusing on the different operations that can be performed in  \gls{sa} and in \gls{nsa} deployments. 
We will also provide a performance comparison among different schemes, along with design insights on the most important parameters related to beam management frameworks.
\end{abstract}

\begin{IEEEkeywords}
5G, mmWave, initial access, tracking, 3GPP, NR.
\end{IEEEkeywords}

\glsresetall
\glsunset{nr}
\glsunset{3gpp}

\section{Introduction}

In order to satisfy the expected growth in capacity demand~\cite{osseiran2014scenarios}, the fifth generation of cellular networks (5G) will adopt for the first time communication at mmWaves in a truly mobile access scenario, thanks to the support of frequencies up to 52.6 GHz in 3GPP Release 15 for \gls{nr}~\cite{38300}. In this band, indeed, there are large chunks of untapped spectrum that can be allocated to the cellular radio access to boost the available data rate and increase the network capacity~\cite{rangan2017potentials}. 

\begin{figure*}[t]
	\centering
	\includegraphics[width=.81\textwidth]{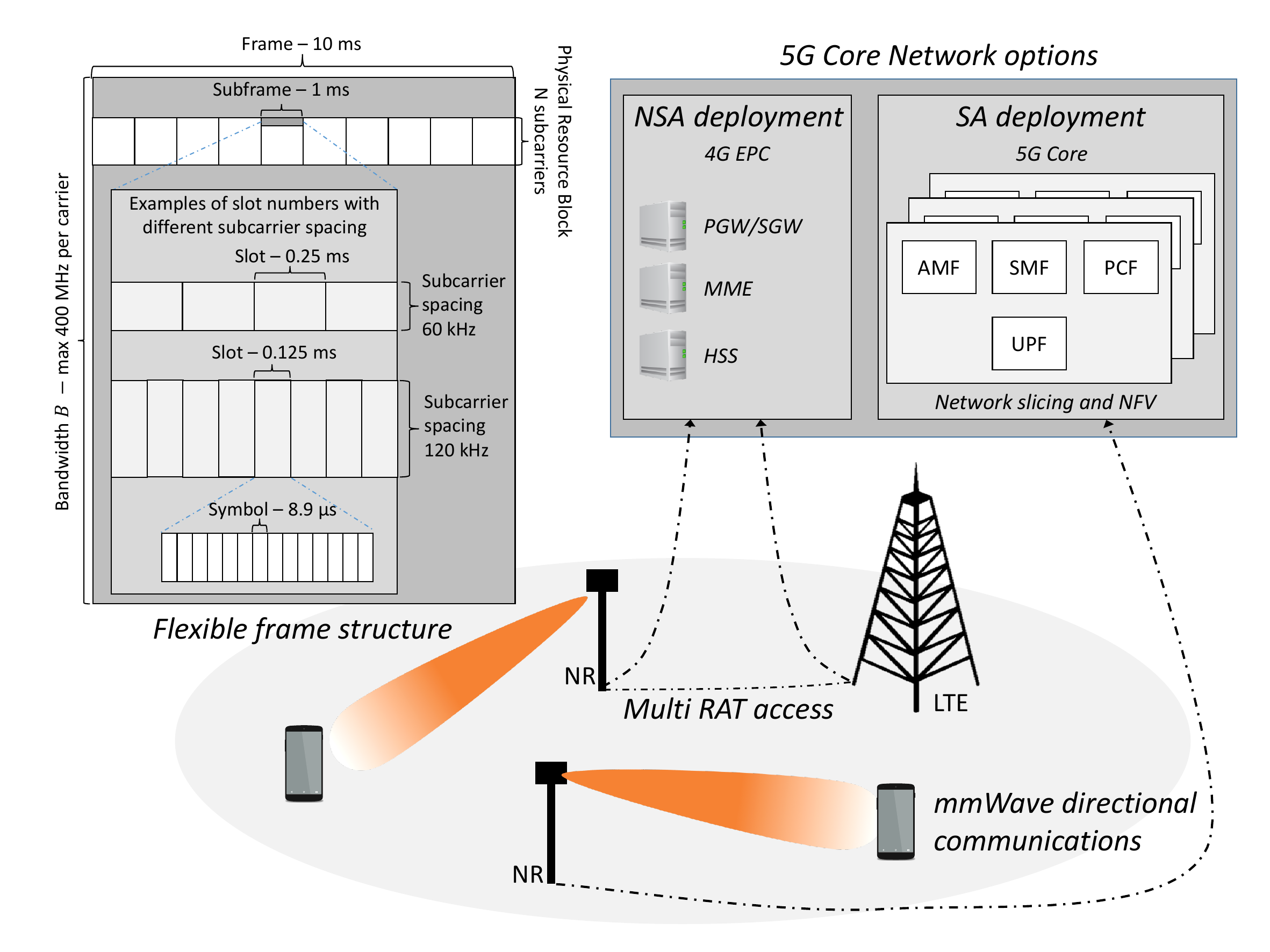}	
	\caption{Key novelties and options of 5G architecture.}
	\label{fig:5gnov}
\end{figure*}

The communication at such high frequencies, however, introduces new challenges for the whole protocol stack, which could affect the end-to-end performance and the quality of experience perceived by the users. The main issues are related to the harsh propagation environment, characterized by high propagation loss and sensitivity to blockage from common materials~\cite{rangan2017potentials}. In order to overcome these problems, the NR standard includes new \gls{phy} and \gls{mac} layer operations, related to the support of directional communications~\cite{38211}, and new deployment options related to multi-connectivity and inter-networking with legacy sub-6 GHz network such as \gls{lte}~\cite{37340}. 

Directional communications are enabled by antenna arrays with a large number of elements, which are feasible at mmWaves thanks to the small wavelength. They provide an additional beamforming gain to the link budget, compensating for the increase in propagation loss, and possibly enabling spatial multiplexing~\cite{rangan2017potentials}. Directional links, however, need accurate alignment of the transmitter and the receiver beams, a procedure which might introduce a delay to access the network and to update the beam pair, thus impacting the overall end-to-end performance. Therefore, reliable networks will need robust and optimized beam management operations, both for \gls{ia}, i.e., when the \gls{ue} is not connected, and for tracking, i.e., when the \gls{ue} is exchanging data with the network.

Multi-connectivity solutions have also been shown to improve the end-to-end performance in mmWave networks, by combining a reliable sub-6 GHz link (e.g., on \gls{lte}) with a high capacity mmWave connection~\cite{polese2017jsac}. Additionally, the \gls{nsa} deployment option~\cite{37340}, in which \gls{lte} and \gls{epc} are used as a radio overlay and as core network also for NR base stations, has been introduced to ease the deployment of NR.

In this article, we will discuss beam management frameworks for NR at mmWave frequencies, focusing in particular on the benefits that multi-connectivity may  introduce. We will compare the operations supported by NR \gls{sa} and \gls{nsa} deployments at mmWaves, and their performance. We show that exploiting sub-6 GHz frequencies reduces the latency in beam reporting during \gls{ia} and in the reaction to \gls{rlf}. On the other hand, thanks to the robust signaling structure designed for NR, the \gls{sa} option exhibits a good performance for \gls{ia}, even though it is more dependent than  \gls{nsa} on the network and antenna array configurations\footnote{For a more technical discussion and an extensive set of results we refer the interested reader to~\cite{giordani2018tutorial}.}. 
In this context, some recent  works (e.g., \cite{liu2018initial,onggosanusi2018modular}) have  provided an overview of the key features pertaining to \gls{ia} and beam management for 5G  NR networks currently being standardized by the 3GPP. 
In \cite{liu2018initial}, the authors focus on a user-centric coordinated transmission mechanism able  to realize seamless intra and inter-cell handover and \gls{ia} operations, and to reduce the interference levels across the network. 
In \cite{onggosanusi2018modular} the use of analog and digital beamforming architectures is investigated as a means to support high-resolution channel state information exchange to deliver efficient beam management procedures.
Our work is distinguished from the aforementioned references as we now investigate, in a single contribution,  the performance and the architectural  implementation of beam management for users both in idle and in connected mode, to enable \gls{ia} and tracking operations, respectively. 
Moreover, unlike previous  papers, this is the first contribution in which both \gls{sa} and \gls{nsa} deployments are considered and compared for beam~management.

The remainder of the paper is organized as follows. In Sec.~\ref{sec:nr} we provide a brief overview on the main NR novelties, and in Sec.~\ref{sec:beams} we describe the NR beam management operations. In Sec.~\ref{sec:results} we report a performance evaluation, and discuss the influence of beam management configurations on the overall results. Finally, in Sec.~\ref{sec:conclusions} we conclude the paper and suggest possible avenues for future work.  

\section{3GPP NR}\label{sec:nr}

\gls{nr} and \gls{5gc} are, respectively, the 3GPP standards for the \gls{ran} and the core network of 5G networks. Their main characteristic is  flexibility: the standard, indeed, provides a general technology framework designed to address the different and, in some cases, conflicting 5G requirements~\cite{osseiran2014scenarios} and to be forward compatible, so that it can accommodate future applications and use cases.
Therefore, the main novelties with respect to \gls{lte} in the \gls{ran} are (i) a more flexible frame structure; (ii) the support of a much larger spectrum, with frequencies also in the mmWave band (up to 52.6 GHz); and (iii) the design of \gls{phy} and \gls{mac} layer procedures for beam management~\cite{38211,38300}, which we will describe in Sec.~\ref{sec:beams}.
In the core network, instead, the new \gls{5gc} introduces network slicing and a higher level of flexibility and virtualization with respect to the traditional \gls{lte} \gls{epc}. Moreover, different deployment options and inter-networking with \gls{lte} are supported.

Fig.~\ref{fig:5gnov} summarizes the main 5G novelties for mobile cellular networks and, in the following paragraphs, we will introduce the main novelties related to the \gls{nr} frame structure and the 5G deployment architectures, focusing on how they can support mmWave communications.

\subsection{\gls{nr} Frame Structure}
\label{sec:frame}

The main characteristics of the frame structure can be found in~\cite{38211}. The waveform will be \gls{ofdm} with a cyclic prefix, and, in general, the frame structure is based on time and frequency sub-divisions similar to those of \gls{lte}, with a frame of 10 ms composed of 10 subframes. However, the main novelty with respect to \gls{lte} is  flexibility: \gls{nr} supports multiple numerologies, i.e., sets of parameters for the \gls{ofdm} waveform, also multiplexed in time and frequency, as long as they are aligned on a subframe basis. This makes it possible to target the different 5G use cases: for example, a shorter \gls{ofdm} symbol duration, combined with a higher subcarrier spacing, can be used for high-data-rate and low-latency traffic, while lower subcarrier spacing can be used for low-frequency narrowband communications for machine-generated traffic.

The top-left corner of Figure~\ref{fig:5gnov} shows an example of \gls{nr} frame structure. The maximum bandwidth for a single carrier is 400 MHz, and it is possible to aggregate up to 16 carriers. In time, each subframe is composed by $2^n$ slots, with $n$ ranging from 0 to 4. This parameter controls also the subcarrier spacing, which is given by $15\times 2^n$ kHz. For frequencies above 6 GHz, the minimum value of $n$ is 2, therefore the minimum subcarrier spacing is 60 kHz. In time, each slot contains 14 \gls{ofdm} symbols, but also mini-slots are supported: they can last as little as 2 \gls{ofdm} symbols, have variable length, and can be positioned asynchronously with respect to the beginning of a standard slot. They are designed for ultra-low latency communications, so that the data transmission does not need to wait for the beginning of the next slot. 

Another important novelty is that a subframe can be self-contained, meaning that a complete round-trip transmission with a first transmission and the corresponding acknowledgment can happen in a single subframe. Therefore, \gls{nr} supports  sub-ms latency for acknowledged transmissions.

\begin{figure*}[t!]
	\centering
	\includegraphics[width=1\textwidth]{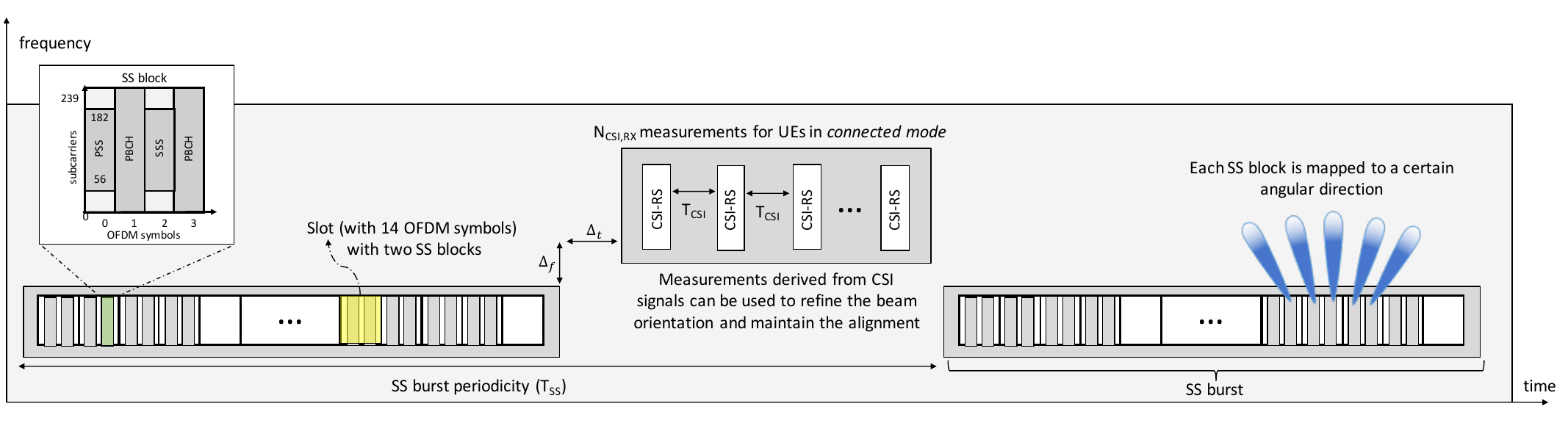}
	\caption{ Example of downlink measurement signal configuration for \gls{nr} systems.  SS blocks are sent every $T_{\rm SS}$, and they embed time and frequency offsets ($\Delta_t$ and $\Delta_f$, respectively) indicating the time and frequency allocation of CSI-RS signals within the frame structure. CSI-RS signals are sent every $T_{\rm CSI}$.  }
	\label{fig:ss_strucutre}
\end{figure*}

\subsection{5G Network Deployment}
\label{sec:deploy}
The flexibility provided by the 3GPP specifications extends to the possible deployment architectures and interconnectivity between 4G and new 5G networks, as shown in Fig~\ref{fig:5gnov}. In particular, in order to smooth the transition between the different generations and reuse the widely deployed \gls{lte} and \gls{epc} infrastructure, the \gls{nr} specifications foresee a \gls{nsa} deployment, in which the network operator only deploys \gls{nr} \glspl{gnb} which are connected to \gls{epc}, possibly with a \gls{dc} setup aided by \gls{lte}. The other option is an \gls{sa} deployment, in which both the \gls{ran} and the core network respect \gls{5g} specifications.
When it comes to deployments at mmWave frequencies, the \gls{nsa} option with \gls{dc} across different \glspl{rat} can also be particularly beneficial to improve the network performance, as shown in~\cite{polese2017jsac}. The so-called \gls{endc}~\cite{37340} is an extension of the \gls{dc} already standardized in \gls{lte} networks, and allows a \gls{ue} to transmit and receive data from different base stations, belonging to an \gls{lte} and an \gls{nr} deployment, with independent schedulers, possibly connected by a non-ideal (i.e., with limited data rate and additional latency) backhaul link. In particular, with \gls{endc} the core network is the \gls{epc}, but 3GPP will standardize also a multi \gls{rat} \gls{dc} option with \gls{5gc}. There is the possibility of performing inter-\gls{rat} measurement and coordination of the user mobility in the different \glspl{rat}, and to use service radio bearers for control only on \gls{lte}, so that the mobility management at mmWave frequencies can be aided by the control plane at lower frequencies. 

\section{Beam Management in 3GPP NR}\label{sec:beams}

MmWave communication systems typically implement directional transmissions, e.g., using high-dimensional phased arrays, to benefit from the resulting beamforming gain and compensate for the increased path loss experienced at high frequencies.
Therefore, next-generation cellular networks must provide a mechanism by which \glspl{ue} and \glspl{gnb} regularly identify the optimal beams to interconnect at any given time.
To this goal, the \gls{3gpp} has specified a set of basic procedures for the control of multiple beams at \gls{nr} frequencies above 6 GHz which are categorized under the term \emph{beam management}~\cite{giordani2018tutorial}. In line with the \gls{3gpp} design for \gls{nr}, we consider a \emph{downlink} architecture where the synchronization and reference signals (i.e., \gls{ss} blocks and \glspl{csirs}, respectively) are broadcast by the \glspl{gnb} and received by the \glspl{ue} within reach. 
 In particular, the following four operations are defined:
\begin{itemize}
 	\item \textit{Beam sweeping}, i.e., the covering of a spatial area with a set of beams transmitted and received according to pre-specified intervals and directions.
 	\item \textit{Beam measurement}, i.e., the evaluation of the quality  of the received signal, which can be expressed in terms of \gls{rsrp} -- the linear average of the received power on different resources with synchronization signals, the \gls{rsrq} -- the ratio between the \gls{rsrp} and the \gls{rssi}, a measurement that includes also thermal noise and signals from other sources, or the \gls{sinr}~\cite{38215}.
 	\item \textit{Beam determination}, i.e., the selection of the optimal beam (or set of beams) to set up a directional (and fully beamformed) communication. 
 	\item \textit{Beam reporting}, i.e., the procedure with which the nodes send to the \gls{ran} information on the quality of the received beamformed signals and on their decision in the beam determination phase. 
 \end{itemize}
For \emph{idle} users, beam management is fundamental to design a directional \emph{initial access} strategy, which allows the \gls{ue} to establish a physical link connection when it first accesses the network~\cite{giordani2016initial,giordani2018tutorial}.
For users in \emph{connected mode}, as the dynamics of the mmWave channel imply that the directional path to any cell can deteriorate rapidly, beam management is required to maintain precise alignment of the transmitter and receiver beams as the \glspl{ue} move, an operation that resembles handover and which is referred to as \emph{tracking}~\cite{polese2017jsac}.

In this context, after reviewing in Sec.~\ref{ssec:signals} the most relevant measurement signals supported by \gls{3gpp} \gls{nr} for beam management, in Sec.~\ref{ssec:MBprocedures} we present \gls{sa} and \gls{nsa} operations for both \gls{ia} and tracking purposes.

\subsection{Downlink Measurement Signals for Beam Management}
\label{ssec:signals}


The structure of the measurement signals considered in the following paragraphs is graphically represented in Fig.~\ref{fig:ss_strucutre}. \\

\textbf{SS Blocks.} In \gls{lte} systems, the synchronization procedure relies on two specifically designed physical signals, which are broadcast omnidirectionally in the downlink, namely the \gls{pss} and the \gls{sss}. Each \gls{ue} in the cell is aware a priori of when and where the synchronization control channel is and can extract and detect those signals.

Along these lines, the \gls{3gpp} has defined a directional version of such signals introducing the concept of \gls{ss} blocks and bursts~\cite{38211}.
An \gls{ss} block is a group of 4 \gls{ofdm} symbols~\cite{38211} in time and 240 subcarriers in frequency (i.e., 20 resource blocks) with the \gls{pss}, the \gls{sss} and the \gls{pbch} that can be used to estimate the \gls{rsrp} and select the optimal beam to communicate. 
The \gls{ss} blocks are grouped into the first 5 ms of an \gls{ss} burst, which can have different periodicities $T_{\rm SS} \in \{5, 10, 20, 40, 80, 160\}$ ms~\cite{38331}. The maximum number $L$ of \gls{ss} blocks in a burst is frequency-dependent, and above 6 GHz there could be up to 64
blocks per burst.

When considering frequencies that need directional transmissions, each \gls{gnb} transmits directionally the \gls{ss} blocks, by sequentially sweeping different angular directions to cover a whole cell sector. 
Based on the measured quality of the received signal, the SS blocks can be exploited by both idle users to identify their initial directions of transmission (in this case, to reduce the impact of SS transmissions and guarantee prompt network access operations, SS can be sent through wide beams) and connected users for beam tracking purposes.

\begin{figure*}[t!]
	\centering
	\includegraphics[width=0.9\textwidth]{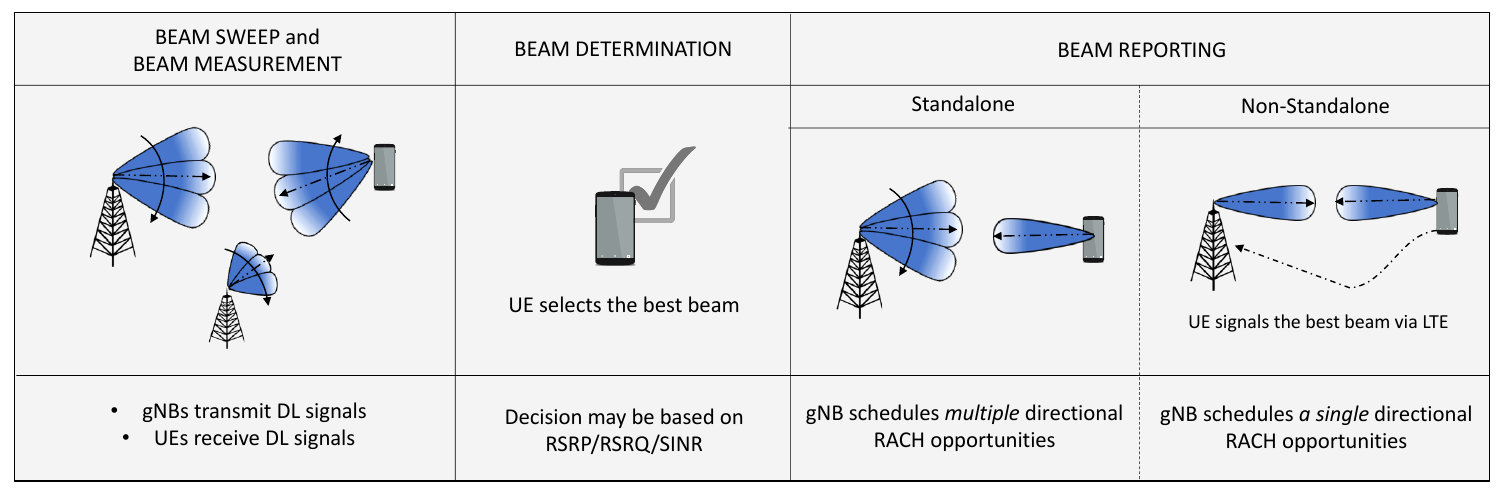}
	\caption{ Graphical representation of the beam management procedures described in Sec.~\ref{ssec:MBprocedures}. \gls{sa} and \gls{nsa} architectures are considered.  }
	\label{fig:beam_management_schemes}
\end{figure*}


\textbf{CSI-RS.}
In \gls{lte}, \glspl{csirs} allow connected \glspl{ue} to regularly estimate the channel conditions and report \gls{cqi} to their serving base station.
Likewise, in \gls{nr}, these signals can be used as \gls{rrm} measurements for mobility management purposes in connected mode~\cite{38300}. 
However, it is fundamental for the \glspl{ue} to know in which time and frequency resources the \glspl{csirs} signals will be sent:
as long as each CSI-RS is represented by a unique identifier, it shall be possible to configure multiple \glspl{csirs} to the same \gls{ss} burst, in such a way that the UE can first obtain synchronization with a given cell using the \gls{ss} bursts, and then use that as a reference to search for CSI-RS resources. 
The \glspl{csirs} configuration should also contain at least the periodicity and time/frequency offsets relative to the associated \gls{ss} burst.

Regarding the bandwidth, although an \gls{nr} network may be able to transmit CSI-RS measurements in full bandwidth, in some deployment scenarios  the reference signals may be broadcast through a subset of all the available frequency resources (with a minimum of 50 resource blocks \cite{38214}) which is deemed sufficiently large to allow proper channel estimation at the receiver.

Regarding the time allocation, \glspl{csirs} may span $N = $1, 2 or 4 \gls{ofdm} symbols~\cite{38211}.
Moreover, the \gls{3gpp} defines different activation methodologies for the \gls{csirs} measures. 
For \emph{periodic} or \emph{semi-persistent} \gls{csirs} transmissions, the following  periodicities (measured in slots) are supported: $T_{\rm CSI} \in \{5, 10, 20, 40, 80,160,320,640\} $\cite{38211}.
For \emph{semi-persistent} and \emph{aperiodic} \gls{csirs} transmissions, the resources are configured and pre-allocated by the higher layers, while their persistent counterparts assume that measurement signals are broadcast with regularity.
For \emph{aperiodic} transmissions, a single set of \gls{csirs} triggering states is higher-layer configured, therefore a \gls{ue} is not expected to receive more than one aperiodic \gls{csirs} in a given slot, i.e., it is not expected to transmit more than one aperiodic \gls{cqi} report to its serving cell~\cite{38214}.
It should be noted that \glspl{csirs} may have a significantly higher time/frequency density compared to that of the \gls{ss} blocks, thus implying higher overhead but, at the same time, higher flexibility.

When considering beamformed communications, the CSI-based results assume a crucial role if they are properly associated with the \gls{ss} measurements.
In this approach, as far as the transceiver has already identified a suitable beam pair to establish a directional transmission, the measurements derived from the \gls{csirs} signals corresponding to different angular directions can be used to refine the beam orientation and maintain the alignment between the communication nodes when considering network topology changes and dynamic evolution of the channel.

\subsection{Standalone vs. Non-Standalone Beam Management}
\label{ssec:MBprocedures}
In this subsection, we focus on the differences between  SA and  NSA deployments for the beam management of users both in idle and in connected mode, as illustrated in Fig.~\ref{fig:beam_management_schemes}.

\textbf{Idle Mode.} Referring to the list of operations described in Sec.~\ref{sec:beams} for beam management, we claim that the \gls{nsa} functionality 
enables an improvement in the \emph{beam reporting} phase, which allows the UE to disseminate the  beam quality and beam decision information to the \gls{ran}.
In order to do so, the mobile terminal has to wait for its candidate serving gNB to schedule a \gls{rach} opportunity towards the best direction it has selected during the  beam determination phase to perform random access. 
If an \gls{sa} deployment is preferred, this may require an additional complete directional scan of the gNB, thus further increasing the time it takes to access the network.\footnote{It has been agreed that, for each SS block, the gNB will specify one or more RACH opportunities with a certain time and frequency offset and direction, so that the UE knows when to transmit the RACH preamble~\cite{38300}.}
On the other hand, leveraging the support of the LTE overlay offered by the \gls{nsa} architecture, the UE can inform the selected serving infrastructure of the optimal direction (or set of directions) through which it has to steer its beam, in order to be properly aligned, through the legacy LTE connection. A single \gls{rach} opportunity with full beamforming capabilities can therefore be immediately scheduled for that direction, without the need to wait for an additional beam sweep at the gNB side.


\textbf{Connected Mode.} 
When the quality of an associated control channel falls below a predefined threshold, i.e., in the case of \gls{rlf}, mechanisms to recover acceptable communication capabilities (e.g., by updating the steering direction of the network nodes or, as a last resort, by handing over to a more robust gNB) need to be quickly triggered upon notifying the network. 
We assess that faster and more efficient \gls{rlf} recovery operations are guaranteed if an  \gls{nsa} architecture is preferred over an \gls{sa} one.
As soon as an impairment is detected, in the case of SA deployments the UE may not be able to properly inform its serving gNB  since the optimal directional path connecting the endpoints is affected by the failure. 
As a consequence, a recovery phase can only be triggered when a new beam configuration is determined, i.e., after the completion of an IA operation. This  may take up to several tens of milliseconds.
 Conversely,  if the failure notification is forwarded through the robust LTE overlay (i.e., by implementing an NSA-based measurement framework), the RLF recovery delay is equal to the latency of a traditional LTE connection, which may be significantly lower than the time it takes to perform IA. The LTE radio  may also serve the UE’s traffic requests until the mmWave directional communication is successfully restored, thereby offering continuous connectivity even during link failures.

The performance of the SA and NSA architectures will be numerically assessed and compared in Sec.~\ref{sec:results}.

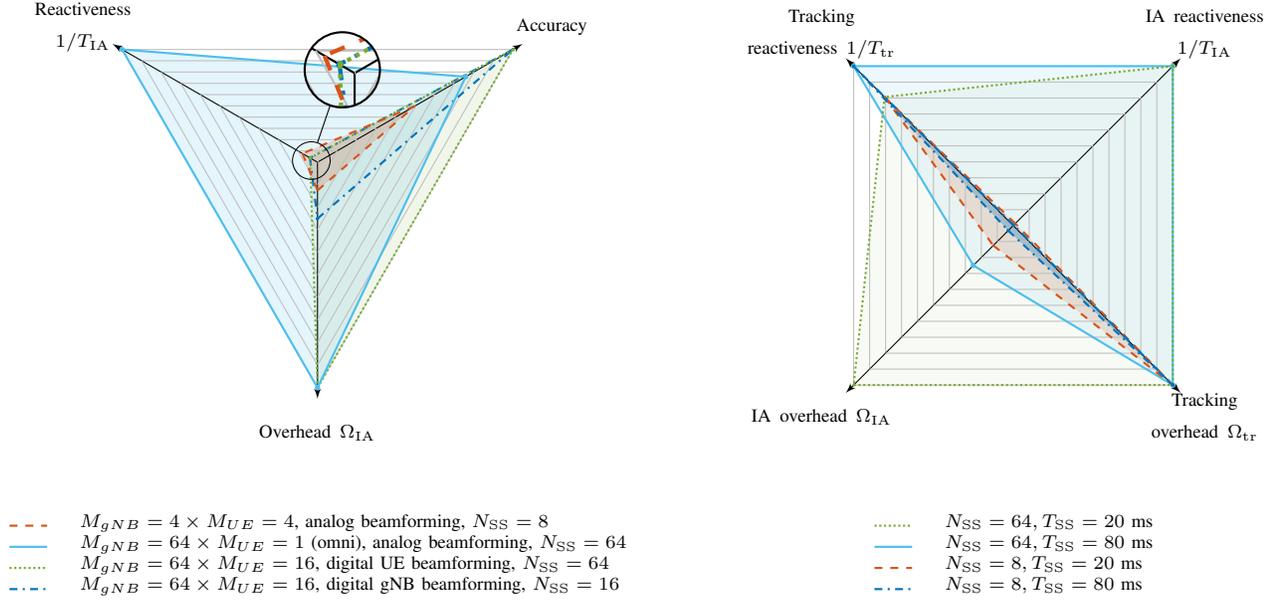
\begin{figure*}
\centering
\setlength\belowcaptionskip{-.15cm}
\begin{subfigure}[t]{0.49\textwidth}
\centering
	\definecolor{mycolor1}{rgb}{0.00000,0.44700,0.74100}%
\definecolor{mycolor2}{rgb}{0.85000,0.32500,0.09800}%
\definecolor{mycolor3}{rgb}{0.92900,0.69400,0.12500}%
\definecolor{mycolor4}{rgb}{0.49400,0.18400,0.55600}%
\definecolor{mycolor5}{rgb}{0.46600,0.67400,0.18800}%
\definecolor{mycolor6}{rgb}{0.30100,0.74500,0.93300}
\pgfplotsset{
tick label style={font=\scriptsize},
label style={font=\scriptsize},
legend  style={font=\scriptsize}
}
\begin{tikzpicture}[spy using outlines=
	{circle, magnification=2, connect spies}, rotate=30]
\tkzKiviatDiagram[scale=0.3,label distance=.9cm,
        radial  = 3,
        gap     = 1,  
        lattice = 10,
        label space  = 2]{\scriptsize Accuracy, \scriptsize Reactiveness $1/T_{\rm IA}$, \scriptsize Overhead $\Omega_{\rm IA}$}

\tkzKiviatLine[dashdotted,thick,mark=x,
                 mark size=2pt,color=mycolor1,fill=mycolor1,opacity=0.05](10, 0.3855, 2.5);\label{6416enbDigiUeAn16} 
\tkzKiviatLine[densely dotted,thick,,color=mycolor5,mark=asterisk,
               fill=mycolor5,opacity=.1](10, 0.3819, 10);\label{6416enbAnUeDig64}
\tkzKiviatLine[thick,color=mycolor6,mark=o,
               fill=mycolor6,opacity=.15](7.5844, 10.0, 10);\label{641enbAnUeAn64}
\tkzKiviatLine[dashed,thick,color=mycolor2,
               fill=mycolor2,opacity=.2](4.97930850353758, 0.7723, 1.25);\label{44ana8}

\coordinate (spypoint) at (-0.2, 0.2);
\coordinate (magnifyglass) at (3,3);

\spy [black, size=1cm] on (spypoint)
   in node at (magnifyglass);

\node[anchor=north,xshift=0pt,yshift=-20pt] at (current bounding box.south) 
{
\scriptsize
\begin{tabular}{@{}ll@{}} 
\tikz\draw[line, thick, color=mycolor2, dashed] (0,1) -- (0.5,1); & $M_{\gls{gnb}} = 4 \times M_{\gls{ue}} = 4$, analog beamforming, $N_{\rm SS}=8$ \\
\tikz\draw[line, thick, color=mycolor6] (0,1) -- (0.5,1);& $M_{\gls{gnb}} = 64 \times M_{\gls{ue}} = 1$ (omni), analog beamforming, $N_{\rm SS}=64$ \\
\tikz\draw[line, thick, color=mycolor5, densely dotted] (0,1) -- (0.5,1); & $M_{\gls{gnb}} = 64 \times M_{\gls{ue}} = 16$, digital UE beamforming, $N_{\rm SS}=64$ \\
\tikz\draw[line, thick, color=mycolor1, dashdotted] (0,1) -- (0.5,1); & $M_{\gls{gnb}} = 64 \times M_{\gls{ue}} = 16$, digital gNB beamforming, $N_{\rm SS}=16$ \\
\end{tabular}
};
\end{tikzpicture}
	\caption{Performance of different directional \gls{ia} schemes with different antenna configurations, i.e., number of antennas at the \gls{gnb} $M_{\gls{gnb}}$ and at the \gls{ue} $M_{\gls{ue}}$, beamforming architecture and number of \gls{ss} blocks per burst $N_{\rm SS}$. The periodicity of the \gls{ss} burst $T_{\rm SS} = 20$ ms is fixed.}
	\label{fig:k1}
\end{subfigure}
\hfill
\begin{subfigure}[t]{0.49\textwidth}
\centering
	\definecolor{mycolor1}{rgb}{0.00000,0.44700,0.74100}%
\definecolor{mycolor2}{rgb}{0.85000,0.32500,0.09800}%
\definecolor{mycolor3}{rgb}{0.92900,0.69400,0.12500}%
\definecolor{mycolor4}{rgb}{0.49400,0.18400,0.55600}%
\definecolor{mycolor5}{rgb}{0.46600,0.67400,0.18800}%
\definecolor{mycolor6}{rgb}{0.30100,0.74500,0.93300}
\pgfplotsset{
tick label style={font=\scriptsize},
label style={font=\scriptsize},
legend  style={font=\scriptsize}
}
\begin{tikzpicture}[spy using outlines=
	{circle, magnification=1.5, connect spies}, rotate=45]
\tkzKiviatDiagram[scale=0.3,label distance=.5cm,
        radial  = 3,
        gap     = 1,  
        lattice = 10,
        label space  = 2]{\scriptsize IA reactiveness $1/T_{\rm IA}$, \scriptsize Tracking reactiveness $1/T_{\rm tr}$, \scriptsize IA overhead $\Omega_{\rm IA}$, \scriptsize Tracking overhead $\Omega_{\rm tr}$}

\tkzKiviatLine[thick,color=mycolor6,mark=o,
               fill=mycolor6,opacity=.1](10.0000,   10.0000,    2.5000,   10.0000
);\label{nss64tss80}

\tkzKiviatLine[densely dotted,thick,color=mycolor5,mark=asterisk,
               fill=mycolor5,opacity=.05](10.0000,    8.0645,   10.0000,   10.0000);\label{nss64tss20}

\tkzKiviatLine[dashed,thick,color=mycolor2,
               fill=mycolor2,opacity=.15](0.1303,    8.0645,    1.2500,   10.0000
);\label{nss8tss20}

\tkzKiviatLine[dashdotted,thick,mark=x,
             	mark size=2pt,color=mycolor1,
             	fill=mycolor1,opacity=0.3](0.0326,   10.0000,    0.3125,   10.0000
);\label{nss8tss80} 


\node[anchor=north,xshift=0pt,yshift=-20pt] at (current bounding box.south) 
{
\scriptsize
\begin{tabular}{@{}ll@{}} 
\tikz\draw[line, thick, color=mycolor5, densely dotted] (0,1) -- (0.5,1); & $N_{\rm SS}=64, T_{\rm SS}=20$~ms \\
\tikz\draw[line, thick, color=mycolor6] (0,1) -- (0.5,1); & $N_{\rm SS}=64, T_{\rm SS}=80$~ms \\
\tikz\draw[line, thick, color=mycolor2, dashed] (0,1) -- (0.5,1);& $N_{\rm SS}=8, T_{\rm SS}=20$~ms \\
\tikz\draw[line, thick, color=mycolor1, dashdotted] (0,1) -- (0.5,1);& $N_{\rm SS}=8, T_{\rm SS}=80$~ms \\
\end{tabular}
};
\end{tikzpicture}
	\caption{Performance of different directional \gls{ia} schemes with different number of \gls{ss} blocks per burst $N_{\rm SS}$ and \gls{ss} burst periodicities $T_{\rm SS}$. The antenna configuration is fixed, with digital beamforming at the \gls{gnb} and analog at the \gls{ue}, with 64 and 16 antenna elements respectively.}
	\label{fig:k2}
\end{subfigure}
\caption{Performance of beam management frameworks for the \gls{ia} and the tracking as a function of different parameters. The metrics considered represent the \textit{accuracy} of the framework (which is inversely proportional to the misdetection probability), the \textit{reactiveness} (which is inversely proportional to $T_{\rm IA}$ and $T_{\rm tr}$) and the overhead.}
\label{fig:k}
\end{figure*}
 
\section{Design Considerations and Results}
\label{sec:results}

In this section we  provide some general guidelines for the selection of the beam management  parameters described in Sec.~\ref{sec:beams}, and we compare the performance of the SA and NSA architectures for beam reporting and RLF recovery operations. We refer to~\cite{giordani2018tutorial} for a complete analysis of the different beam management schemes for \gls{nr}.

\textbf{General Discussion.}
The most important metrics to be considered when evaluating the performance of the beam management schemes are detection accuracy, reactiveness and overhead, as shown in  Fig.~\ref{fig:k}.
The accuracy represents the capability of the framework to identify and correctly measure the beams, and is inversely proportional to the probability of misdetection, i.e., the probability of not detecting the beam. The reactiveness, instead, describes how quickly the framework is able to detect an updated channel condition, and is inversely proportional to the average time $T_{\rm IA}$ required to find the best beam pair, during the \gls{ia}, and to the average time $T_{\rm tr}$ to receive a \gls{csirs} in a certain direction for the tracking of already connected users. Finally, the overhead is the ratio between the number of time and frequency resources that should be allocated to beam management operations (instead of data transmission) and all the available resources. 
Although these metrics are apparently orthogonal there exist some configurations that are able to provide performance gains on different metrics, jointly.

In this context, the beamforming architectures and  the number of antennas at the \glspl{gnb} and the \glspl{ue} (i.e., $M_{gNB}$ and $M_{UE}$, respectively) are key parameters in the design of directional initial access and tracking. 
From Fig.~\ref{fig:k1}, we observe that a larger number of antennas enables narrower beams which, in turn, guarantee better accuracy (thanks to the higher gains achieved by beamforming). On the other hand, highly directional communications lead to worse performance in terms of reactiveness and overhead (due to the increased number of directions that need to be scanned before the optimal beam configuration is selected).
A digital beamforming architecture (which allows the processing of the received signals in the digital domain, enabling the transceiver to generate beams in multiple directions at the same time) has the potential to improve the reactiveness of the measurement scheme and decrease the overhead, without penalizing the accuracy. However, it suffers from increased power consumption with respect to an analog~strategy.\footnote{For completeness, it should be mentioned that the 5G equipment manufacturers are also considering a hybrid beamforming solution, which uses $K_{\rm BF}$ radio-frequency chains and enables the transceiver to transmit/receive in $K_{\rm BF}$ directions simultaneously. Nevertheless, when hybrid beamforming is used for transmission, the power available at each transmitting beam is the total node power constraint divided by $K_{\rm BF}$, thus potentially reducing the received power.}

The setup of the maximum number of SS blocks in a burst (i.e., $N_{SS}$) and the SS burst periodicity (i.e., $T_{SS}$) has also a remarkable impact in terms of reactiveness and overhead, as represented in Fig.~\ref{fig:k2}. 
A higher number of SS blocks per burst, although increasing the overhead linearly, increases the probability of completing the sweep in a single burst  and thus reduces the time it takes to perform IA. 
In these circumstances, a higher $T_{SS}$ would  guarantee more reactive tracking operations and reduce the overhead of the SS blocks, as shown by the \textit{``IA overhead''} axis in Fig.~\ref{fig:k2}.

\begin{figure*}[t]
\centering
\begin{subfigure}[t]{0.4\textwidth}
\centering
\setlength\fwidth{0.65\columnwidth}
\setlength\fheight{0.55\columnwidth}
%
%
\definecolor{mycolor1}{rgb}{0.20810,0.16630,0.52920}%
\definecolor{mycolor2}{rgb}{0.97630,0.98310,0.05380}%
\definecolor{mycolor3}{rgb}{0.00000,0.70000,0.70000}%

\pgfplotsset{
tick label style={font=\scriptsize},
label style={font=\scriptsize},
legend  style={font=\scriptsize}
}
\begin{tikzpicture}

\begin{axis}[
scaled y ticks = true,
width=0.975\fwidth,
height=\fheight,
at={(0\fwidth,0\fheight)},
scale only axis,
bar shift auto,
xmin=0,
xmax=4,
xtick={1,2,3},
xticklabels={4,16,64},
ymode=log,
 log origin=infty,
xlabel style={font=\color{white!15!black}},
xlabel={\footnotesize $M_{gNB}$},
ymin=0,
ylabel style={font=\color{white!15!black}},
ylabel={\footnotesize $T_{\rm BR}$ [ms]},
axis background/.style={fill=white},
xticklabel style={text width=6, align=center}, xmajorgrids,
ymajorgrids,
legend columns=2,
ylabel shift=-8pt,
yticklabel shift=-3pt,
legend style={at={(0.5,1)}, anchor=south, legend cell align=left, align=left, draw=white!15!black,/tikz/every even column/.append style={column sep=0.35cm}},
]
\addplot[ybar, bar width=100, fill=black, opacity=0.1, draw=white, area legend, forget plot] table[row sep=crcr] {%
5 10\\
};
\addplot[color=black, line width=0.5, dashed, forget plot] coordinates {(-6,10) (6,10)};
\node[color=black, font=\scriptsize] at (1,8) {$T_{\rm BR, NSA} = 10$~ms};

\addplot[ybar, bar width=100, fill=blue, opacity=0.1, draw=white, area legend, forget plot] table[row sep=crcr] {%
5 0.8\\
};
\addplot[color=blue, line width=0.5, dashed, forget plot] coordinates {(-6,0.8) (6,0.8)};
\node[color=black, font=\scriptsize] at (1,0.65) {$T_{\rm BR, NSA} = 0.8$~ms};

\addplot[ybar, bar width=100, fill=red, opacity=0.1, draw=white, area legend, forget plot] table[row sep=crcr] {%
5 40\\
};
\addplot[color=red, line width=0.5, dashed, forget plot] coordinates {(-6,40) (6,40)};
\node[color=black, font=\scriptsize] at (1,32) {$T_{\rm BR, NSA} = 40$~ms};

\addplot[ybar, bar width=0.2, fill=mycolor1, draw=black, area legend] table[row sep=crcr] {%
1	0.0625\\
2	0.5\\
3	40.56\\
};
\addlegendentry{$N_{SS}=8$}

\addplot[ybar, bar width=0.2, fill=mycolor2, draw=black, area legend] table[row sep=crcr] {%
1	0.0625\\
2	0.5\\
3	1.562\\
};
\addlegendentry{$N_{SS}=64$}


\end{axis}
\end{tikzpicture}%
\caption{Reactiveness. $T_{SS} = 20$ ms, $\Delta_f = 120$~KHz.\label{fig:hist_BR}}
\end{subfigure}\hfill
\begin{subfigure}[b]{0.6\textwidth}
\centering
\footnotesize
\renewcommand{\arraystretch}{1.2}
\begin{tabular}[b]{lcccc|cc}
\toprule
\multirow{3}{*}{$M_{gNB}$} & \multicolumn{4}{|c|}{Standalone} & \multicolumn{2}{c}{Non Standalone}\\
\cline{2-7}
                             & \multicolumn{2}{|c}{$\Omega_{\rm BR}$ $\cdot10^{-3}$}      & \multicolumn{2}{c|}{$P_C$ [W]} & \multirow{2}{*}{$\Omega_{\rm BR}$ $\cdot10^{-3}$} &  \multirow{2}{*}{$P_C$ [W]}    \\
                             & \multicolumn{1}{|c}{Analog}                  & Digital                & Analog                  & Digital    &             \\ \cline{1-7}
\multicolumn{1}{c|}{4}                             & 0.0894               & 0.0894              &  16.2847               & 64.359 &   0.0894  &    16.2847      \\
\multicolumn{1}{c|}{16}                            &  0.7149              & 0.0894              & 135.8934              & 257.433 &     0.0894  & 16.9867       \\
\multicolumn{1}{c|}{64}                            & 2.2341               & 0.0894              & 494.8670              & 1030.74 &  0.0894  & 19.7947        \\ \bottomrule
\end{tabular}
\caption{Overhead and power consumption.
Non-standalone always requires a single RACH opportunity to perform beam reporting, therefore  analog beamforming (less power consuming than its digital counterpart) is employed. \label{tab:ov_BR}}
\end{subfigure}
\caption{Beam reporting performance considering an SA or NSA architecture. Analog or digital beamforming is implemented at the gNB side, for different gNB antenna array structures and \gls{ss} block configurations. The UE is already steering through its selected direction, therefore beam sweeping is not required.}
\label{fig:beam_reporting}
\end{figure*}

\begin{table*}[t]
\centering
\small
\begin{tabular}{ccclllclcl}
\toprule
\multicolumn{2}{c|}{Antenna}                                                & \multicolumn{8}{c}{$T_{\rm RLF,SA}$ {[}ms{]}}                                                                                                                                                                                                                                                                                                                                                                              \\
\multirow{2}{*}{$M_{gNB}$} & \multicolumn{1}{c|}{\multirow{2}{*}{$M_{UE}$}} & \multicolumn{4}{c}{\multirow{2}{*}{\begin{tabular}[c]{@{}c@{}}$N_{\rm SS}=8$, $T_{\rm SS}=20$\\ $\gls{gnb}$ ABF, UE ABF\end{tabular}}} & \multicolumn{2}{c}{\multirow{2}{*}{\begin{tabular}[c]{@{}c@{}}$N_{\rm SS}=64$, $T_{\rm SS}=40$\\ $\gls{gnb}$ DBF, UE ABF\end{tabular}}} & \multicolumn{2}{c}{\multirow{2}{*}{\begin{tabular}[c]{@{}c@{}}$N_{\rm SS}=64$, $T_{\rm SS}=80$\\ $\gls{gnb}$ DBF, UE ABF\end{tabular}}} \\
                           & \multicolumn{1}{c|}{}                          & \multicolumn{4}{c}{}                                                                                                                   & \multicolumn{2}{c}{}                                                                                                                    & \multicolumn{2}{c}{}                                                                                                                    \\ \midrule
4                          & \multicolumn{1}{c|}{4}                         & \multicolumn{4}{c}{30.2322}                                                                                                              & \multicolumn{2}{c}{20.3572}                                                                                                                   & \multicolumn{2}{c}{40.3572}                                                                                                                   \\
64                         & \multicolumn{1}{c|}{1}                         & \multicolumn{4}{c}{130.1072}                                                                                                              & \multicolumn{2}{c}{20.0535}                                                                                                                   & \multicolumn{2}{c}{40.0535}                                                                                                                   \\
64                         & \multicolumn{1}{c|}{16}                        & \multicolumn{4}{c}{ 5250}                                                                                                              & \multicolumn{2}{c}{22.6072}                                                                                                                   & \multicolumn{2}{c}{42.6072}                                                                                                                   \\ \midrule
\multicolumn{10}{c}{$T_{\rm RLF,\gls{nsa}}\in\{ 10, 4, 0.8\}$ ms, according to the considerations in \cite{latencyreduction2017}.}                                                                                                                                                                                                                                                                                                                                                                                                                                                    \\ \bottomrule
\end{tabular}
\caption{RLF recovery delay considering the SA or the \gls{nsa} measurement frameworks, for different values of $N_{\rm SS}$, $T_{\rm SS}$ and for different beamforming configurations. $\Delta_f=120$ kHz. ABF stands for Analog Beamforming, and DBF for Digital.}
\label{tab:RLF}
\end{table*}

\textbf{NSA vs. SA.}
As discussed in Sec.~\ref{ssec:MBprocedures}, the design of an NSA-like framework for beam management may be preferable in many respects.
First,  as illustrated in Fig.~\ref{fig:hist_BR}, although the SA architecture generally presents low beam reporting delays $T_{\rm BR}$, an NSA scheme may guarantee faster reporting operations in case of large antenna arrays at the gNB side, i.e., when configuring very narrow  beams which would inevitably increase the number of directions to scan before a RACH opportunity is scheduled towards the correct random access direction.
For the NSA case, the beam reporting delay $T_{\rm BR, NSA}$ is  equal to the latency of an LTE connection which, assuming no retransmissions are needed, ranges from 10.5 ms to 0.8 ms, according to the latency reduction techniques being implemented~\cite{giordani2018tutorial}.
Notice that the results are independent of $M_{UE}$ since the \gls{ue}
has already selected its optimal steering direction and therefore does not require a beam sweeping operation.

Second, from Fig.~\ref{tab:ov_BR}, we observe that an NSA architecture can reduce the impact of the beam reporting overhead $\Omega_{\rm BR}$  while delivering more power-efficient operations.
For the SA case,  the overhead may increase significantly, especially if analog beamforming is implemented, since the completion of the random access requires  the system to  scan through all directions one by one, thereby requiring the allocation of possibly multiple RACH resources.
On the other hand, an NSA scheme necessitates a single RACH opportunity -- regardless of the beamforming architecture being implemented --  with a total overhead of $0.0894 \cdot 10^{-3}$. 
It should be noticed that, while this overhead may seem small, it represents that of a RACH opportunity for a single user, and not the overall overhead of \gls{ia} procedures~\cite{giordani2018tutorial}. Moreover, we recall that the SA overhead may be reduced as configuring a digital beamforming architecture, which enables the transceiver to direct beams at multiple directions simultaneously, thereby removing the need for a directional scan at the gNB side during random access. However, digital beamforming requires a separate \gls{rf} chain for each antenna element and therefore shows much higher power consumption $P_C$ than if analog beamforming were~preferred.\footnote{The  total power consumption $P_C$ of each beamforming scheme is evaluated according to \cite{abbas2016towards} in which $b = 3$ quantization bits are used by the Analog-to-Digital Converter block.}

Third, Tab.~\ref{tab:RLF} exemplifies how, in connected mode, an NSA implementation offers  reactive recovery operations in the case of radio link failure events.
We observe that, for the SA case, the RLF recovery delay $T_{\rm RLF,SA}$  is quite high for all the investigated settings and is dominated by the IA delay. In some circumstances (e.g., $N_{SS} = 8$, $T_{SS} = 20$ ms, $M_{gNB} = 64,$ $M_{UE} = 16$ and when analog beamforming is implemented), the RLF recovery delay assumes unacceptably high values.
For the NSA case, instead,  the RLF recovery delay $T_{\rm RLF,NSA}$ is equal to the latency of a traditional LTE connection (which depends on the implemented latency reduction technique) which, in general, is remarkably lower than the IA delay.  

\section{Conclusions}
\label{sec:conclusions}

A challenge for the feasibility of 5G cellular  systems operating at mmWaves is the rapid channel dynamics that affect a high-frequency environment and the need to maintain alignment between the communication endpoints.
In this regard,  the design and configuration of efficient initial access and tracking procedures able to periodically identify the optimal beam pair with which a base station and a mobile terminal communicate is of extreme importance.
In this paper, after a general description of the main  parameters and reference signals specified by the 3GPP for NR, we compared the performance of a standalone and non-standalone deployment for the management of the beams of users both in connected and in idle modes.
We showed that an NSA configuration exploiting multi-connectivity offers improved end-to-end performance in mmWave networks and has the potential to (i) guarantee higher resilience and  improved reactiveness in case radio link failures occur, (ii) reduce the impact of the overhead in the beam reporting phase,   and (iii)  deliver more reactive  reporting~operations.

\bibliographystyle{IEEEtran}
\bibliography{bibl.bib}

\end{document}